# Are Ultrathin Stents Optimal for Bifurcation Lesions? Insights from Computational Modelling of Provisional and DK-Crush Techniques

Thin vs Ultrathin Stents in Bifurcation PCI

Andrea Colombo[a], Dario Carbonaro[b], Mingzi Zhang[a], Chi Shen[a], Ramtin Gharleghi[c], Ankush Kapoor[a], Claudio Chiastra[b], Nigel Jepson[d,e], Mark Webster[f], Susann Beier[a]

[a] Sydney Vascular Modelling Group, School of Mechanical and Manufacturing Engineering, University of New South Wales, Sydney, NSW, Australia

[b] Polito[BIO]Med Lab, Department of Mechanical and Aerospace Engineering, Politecnico di Torino, Turin, Italy

[c] Independent researcher

[d] Prince of Wales Clinical School of Medicine, University of New South Wales, Sydney, NSW, Australia

[e] Prince of Wales Hospital, Sydney, NSW, Australia

[f] Green Lane Cardiovascular Service, Auckland City Hospital, Auckland, New Zealand

**Corresponding author:**
Andrea Colombo
School of Mechanical and Manufacturing Engineering, Ainsworth Building (J17), Engineering Rd, University of New South Wales, Kensington, NSW, 2052, Australia
Email: a.colombo@unsw.edu.au
https://orcid.org/0009-0009-6436-6785

**Abbreviations**: CFD: Computational Fluid Dynamics, CT: Computed Tomography, DKC: Double Kissing Crush, DV: Distal Vessel, FEA: Finite Element Analysis, KBI: Kissing Balloon Inflation, MV: Main Vessel, PCI: Percutaneous Coronary Intervention, POT: Proximal Optimization Technique, PSB: Provisional Side Branch, SB: Side Branch, TAESS: Time-Averaged Endothelial Shear Stress.

**Structured Abstract**

**Background**

Complex coronary bifurcation lesions remain challenging in percutaneous coronary intervention, with stent design and deployment strategy influencing clinical outcomes. This study compares the mechanical and hemodynamic performance of the ultrathin-strut Orsiro and thin-strut Xience Sierra stent in Provisional Side Branch (PSB) and Double Kissing Crush (DKC) techniques.

**Methods**

We used finite element analyses of bifurcation stent deployment to assess malapposition, ostium clearance, and arterial wall stress for both techniques. Computational fluid dynamics simulations quantified the luminal exposure to low Time-Averaged Endothelial Shear Stress (TAESS < 0.4 Pa) and high shear rates (>1000 $s^{-1}$).

**Results**

In PSB, Orsiro showed higher malapposition (13.0% vs 9.6%) but improved SB ostium clearance (77% vs 64%) and lower low-TAESS exposure (30.3% vs 33.6%) compared to Xience. Orsiro also produced higher arterial wall stresses, particularly during kissing balloon inflation. In DKC, differences in malapposition and ostium clearance diminished between stents, though Orsiro retained a hemodynamic advantage with lower low-TAESS (28.2% vs 36.3%).

**Conclusions**

Stent design influenced outcomes more strongly in PSB, where anatomical interaction and platform-specific behavior impacted both structural and hemodynamic results. In DKC, procedural complexity minimized those differences, making the stenting technique the primary performance driver. Nonetheless, Orsiro consistently preserved more favorable flow conditions. These findings highlight the need to match device selection with lesion characteristics in PSB, while in DKC, optimizing procedural steps may have a greater impact than the choice of stent platform.

- **Novelty.** First high-fidelity computational model of double kissing crush deployment, allowing a quantitative comparison of stent platforms in a robust, controlled, and repeatable setting.
- **Stent Design Matters in PSB.** In the PSB technique, Xience Sierra showed better strut apposition, ostium circularity, and reduced arterial wall stress, while Orsiro demonstrated overall more favorable hemodynamics, associated with reduced restenosis risk.
- **Stent Design is Less Impactful in DKC.** Under the DKC technique, differences in malapposition, arterial stress, and ostium clearance between Orsiro and Xience Sierra were modest, suggesting that the technique itself dominated overall outcomes. However, Orsiro retained a hemodynamic advantage.
- **Stenting Technique.** PBS vs DKC is a more impactful choice than the stenting platform. Compared to provisional side branch stenting, the multi-step double kissing crush technique increases procedural complexity and leads to higher strut malapposition and worse hemodynamics regardless of the stent platform.
- **Procedural Steps:** Kissing balloon inflation induces the highest arterial stresses, while the proximal optimization technique induces the highest stent stresses. Careful execution of these steps is important in minimizing procedural risks.

## 1. INTRODUCTION

Orsiro (Biotronik AG, Bulach, Switzerland) and Xience Sierra (Abbott Vascular, Abbott Park, IL, USA) stents are widely employed in Percutaneous Coronary Intervention (PCI) due to their potential to reduce restenosis rates and improve vessel healing[1]. Orsiro has an ultrathin-strut design with a thickness of 60 µm, while Xience Sierra is classified as thin-strutted, with a strut thickness of 81 µm. Clinical trials comparing these stents generally found no significant differences in long-term efficacy across a broader patient population[2-4]. Nevertheless, Orsiro has demonstrated favorable outcomes in specific subgroups, such as small vessels, STEMI, and lesions at higher restenosis risk, likely due to its thinner struts and improved healing profile[5-8]. Conversely, Xience Sierra, characterized by its stiffer strut design, may be better suited for complex or heavily calcified lesions requiring greater mechanical support[9-11]. Thus, the current evidence does not conclusively favor one stent over the other for PCI applications[1].

Stent design is a key factor in PCI outcomes, as it directly affects both the mechanical performance of the device and the local hemodynamic (blood flow dynamic) environment[12]. Key design parameters, such as the stent's strut thickness and connector configuration, influence how the stent deforms, interacts with the arterial wall, and modulates blood flow, partly independent of the stent deployment. For example, thicker struts and a higher number of connectors provide increased radial and longitudinal stiffness[13-15]. In contrast, thinner struts and fewer connectors tend to better preserve physiological blood flow conditions and increase flexibility, though often at the expense of reduced radial stiffness[13,16]. These trade-offs underscore the critical need to carefully evaluate stent design characteristics, particularly in bifurcation lesions, where mechanical and blood flow-related factors are amplified.

In addition to stent design, the long-term clinical performance of a stent is also closely influenced by the interventional strategy by which it is deployed[17]. Different bifurcation stenting techniques create distinct local mechanical and hemodynamic outcomes with the same stent device, which may amplify or mitigate its design advantages. For instance, on one hand, the Provisional Side Branch (PSB) technique, a single-stent approach, is generally preferred for non-complex bifurcation lesions[18]. On the other hand, the Double Kissing Crush (DKC) technique, which involves two stents, is recommended in complex bifurcation lesions (as defined by the DEFINITION study criteria, including lesion length, calcification, and bifurcation angle[19]) where maintaining optimal side branch patency is critical[20]. However, while stent design and stenting technique have each been evaluated independently, their combined influence has not been investigated yet. The interplay between device characteristics and procedural strategy may be critical in determining clinical outcomes, particularly in bifurcation lesions, and needs to be studied.
Computational simulations have emerged as a powerful tool for evaluating coronary bifurcation stenting[17] by enabling quantitative analyses of key parameters that directly affect stenting outcome for the first time[17]. These include malapposition, side branch ostium clearance, arterial stress, and key hemodynamic metrics representing the Time-Averaged Endothelial Shear Stress (TAESS)[21-24]. Computational simulations facilitate robust and reproducible comparisons of stent platforms and deployment techniques, which is not possible on the bench, with limited insights, or clinically, with a lack of control and reproducibility. Whilst both Orsiro and Xience Sierra have independently been studied using numerical simulations[21] and bench testing[25], different procedural settings prevent a comparative evaluation of these works and thus the stents' performance. Moreover, while the leading structural mechanics Finite Element Analysis (FEA) method has been used to computationally study stenting techniques, including PSB[23,26-33], TAP[27,28], and culotte[26-28,34,35], it has never been applied to the DKC technique. As a result, the biomechanical behavior of DKC remains poorly understood, and no study to date has combined mechanical and blood flow hemodynamic simulations to assess its performance in a unified framework. Addressing this gap is essential to understanding how leading stent designs and common procedural implant strategies interact in complex bifurcation lesion interventions.
This study employs structural mechanics FEA and Computational Fluid Dynamics (CFD) to compare the ultrathin-strut Orsiro and thin-strut Xience Sierra stents, both in PSB and DKC settings. This work represents the first biomechanical model of the DKC procedure and the first direct computational comparison between these two contemporary stents across both stenting strategies. Ultimately, this analysis aims to provide new insights to delineate stent design and implant procedure interplay to ultimately improve PCI success rates.

## 2. METHODS

### 2.1. Stent Deployment

A population-representative coronary bifurcation model was developed with dimensions derived from a Coronary Computed Tomography Angiography (CCTA) atlas[36] consistent with Finet's law to respect the fractal nature of a coronary bifurcation. The branch diameters are as follows: Main Vessel (MV) 3.25 mm, Distal Vessel (DV) 2.5 mm, and Side Branch (SB) 2.25 mm. Arterial wall thickness and bifurcation angle were 0.89 mm and 70°, respectively[33,36].

Stent geometries were digitally reconstructed using a MATLAB-based builder from micro-CT images of the ultrathin-strut Orsiro stent, characterized by a helical pattern, and the thin-strut Xience Sierra stent, a ring-based design[37-39] (Figure 1). Stent sizes were selected from commercially available options suitable for the target vessel. Specifically, for the Orsiro stents, dimensions were 2.5 mm and 13 mm for the MV, and 2.25 mm and 9 mm for the SB. For the Xience Sierra stents, diameters and lengths were 2.5 mm and 12 mm for the MV, and 2.25 mm and 8 mm for the SB.

Stent deployment was simulated using FEA with Abaqus Explicit (Dassault Systèmes, Providence, USA). Stent crimping was simulated to replicate the pre-stress state of the device before deployment. Subsequently, established clinical deployment procedures were followed to replicate both stenting techniques accurately[40]. The procedural steps are summarized in Figure 2, where differences between the simpler PSB approach and the multi-step DKC technique are illustrated. Key balloon expansions and rewiring phases are highlighted, clarifying the steps modeled and the positioning of each balloon relative to the bifurcation anatomy and previously placed stents. In PSB, side branch rewiring was performed distally through the stent cell closest to the carina, while in DKC, both balloons were rewired proximally through the most accessible cell. Rewiring positioning was achieved in HyperMesh (Altair Engineering, Troy, MI, USA). Proximal Optimization Technique (POT) and Kissing Balloon Inflation (KBI) were applied in both techniques to optimize strut apposition and restore the fractal geometry. More information about the simulation set-up can be found in Supporting Information 1.

**Figure 1** – Geometries of the Orsiro and Xience Sierra stents used in the simulations. Orsiro (top) presents a helical structure with ultrathin struts (60 µm), while Xience Sierra (bottom) features a ring-based design with thin struts (81 µm). The commercially available stent lengths differ for the target vessel, with Orsiro measuring 13 mm and 9 mm, and Xience Sierra measuring 12 mm and 8 mm for *MV* and *SB* stent, respectively. *MV: Main Vessel; SB: Side Branch*.

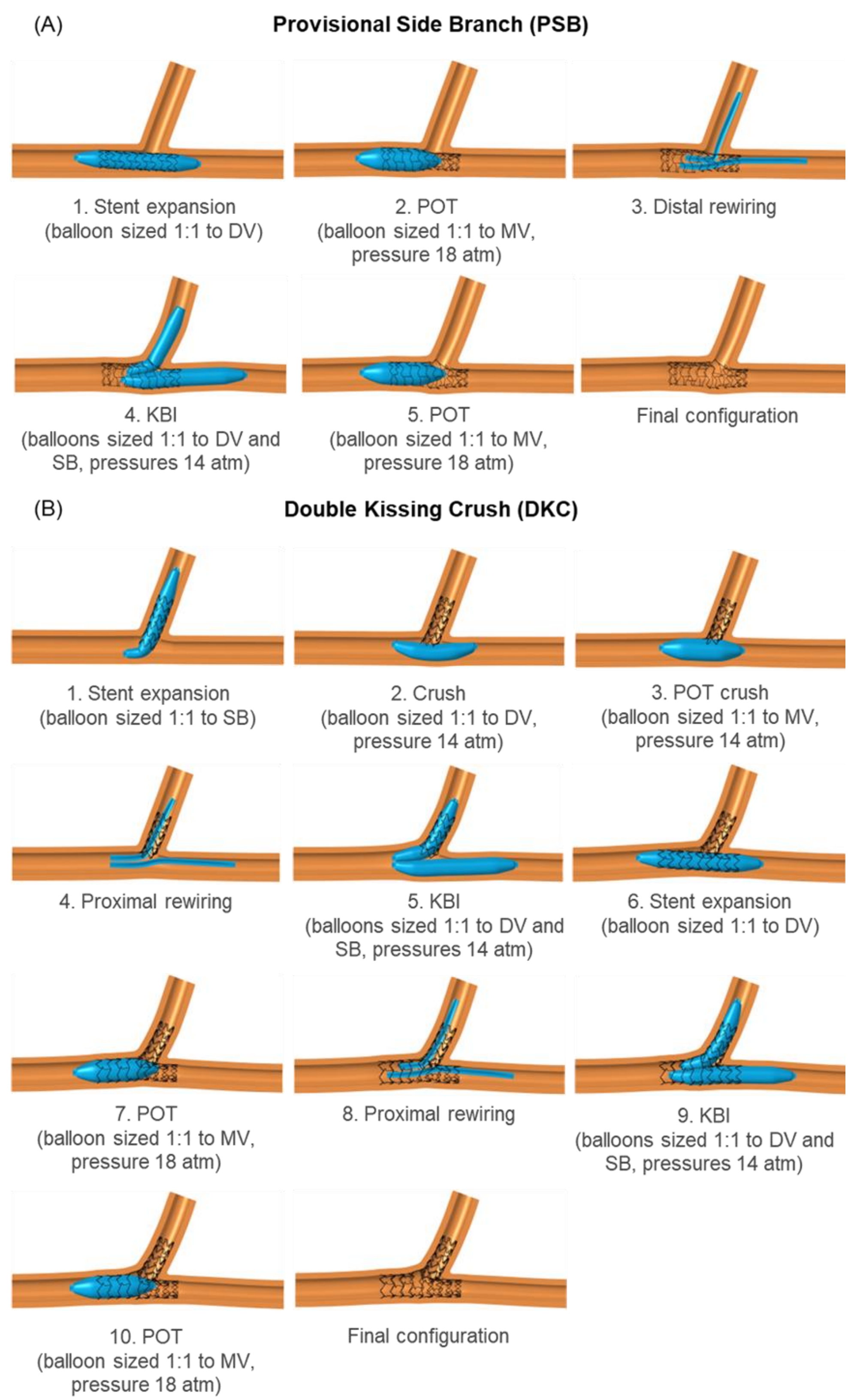

**Figure 2** – Procedural steps of provisional side branch (A) and double kissing crush (B) stenting techniques implemented using structural mechanics finite element simulations. Each panel illustrates a key deployment stage, including balloon sizing and positioning for stent expansion, rewiring steps, POT, KBI, and final stent configurations for both techniques (left to right). *DV: Distal Vessel; KBI: Kissing Balloon Inflation; MV: Main Vessel; POT: Proximal Optimization Technique; SB: Side Branch*.

## 2.2. Blood Flow Simulation

Pulsatile hemodynamic analysis was performed using ANSYS CFX software (ANSYS, Inc., Canonsburg, PA, USA) on the final geometrical configuration after the deployment simulation. The blood was modeled as an incompressible fluid with non-Newtonian viscosity[41,42]. A pulsatile velocity profile, indicative of physiological coronary blood flow, was applied at the MV inlet[16]. The SB outlet had a zero-pressure boundary condition[43], while the DV outlet was assigned a predefined outflow rate based on physiological flow splitting[44]. A mesh and time sensitivity analysis was performed to ensure the reliability of results within a 2% error. For each case, four cardiac cycles were simulated and results extracted from the last cycle only to minimize transient start-up effects (Supporting Information 1).

## 2.3. Stent Performance Evaluation

Stenting performance was assessed using mechanical (FEA-derived) and hemodynamic (CFD-derived) indicators. Mechanical performance was evaluated using malapposition, SB ostium clearance, arterial wall stress, and peak stent stress. Malapposition was defined as the percentage of stent struts located at a distance greater than the strut thickness from the vessel wall over the total strut area. SB ostium clearance was evaluated to measure how well the SB ostium remained unobstructed after stenting, and it was computed as the ratio between the largest open cell area and the total SB ostium area. Arterial wall stress, representing the mechanical stretch within the artery, was calculated by averaging the maximum principal stress values of mesh elements exceeding 10 kPa at the bifurcation region. This threshold was selected in the absence of consensus values in the literature, to exclude regions experiencing negligible stress unlikely to influence tissue response, while still capturing physiologically relevant mechanical loading. In addition, the total volume of arterial tissue experiencing stress above this threshold was quantified to assess the spatial extent of elevated stress. Peak stent stress was separately evaluated to identify potential regions of mechanical failure. Hemodynamic performance was evaluated by calculating TAESS, which represents the mean cardiac-cycle magnitude of shear forces acting on the endothelium lining the lumen throughout the cardiac cycle. In addition, shear rates above 1000 $s^{-1}$ were evaluated due to their association with increased thrombogenic risk[45]. A shear rate burden index was calculated as the product of the blood volume exposed to shear rates > 1000 $s^{-1}$ and the average shear rate within that volume. More details on the metrics analyzed and their clinical significance can be found in Table 1.

**Table 1** - Summary of key mechanical and hemodynamic performance metrics evaluated in this study, along with desired ranges or patterns and associated clinical relevance. These indicators support the interpretation of stent behavior and potential implications for restenosis, thrombosis, and vessel healing. *SB: Side Branch; TAESS: Time-Averaged Endothelial Shear Stress.*

| Metric | Desired Range | Clinical Significance |
|---|---|---|
| Malapposition | To be minimized | Associated with thrombus formation and delayed endothelial healing, potentially increasing restenosis risk[46]. |
| SB ostium clearance | To be maximized | Linked clinically to late stent thrombosis[25,47]. |
| Arterial wall stress | To be minimized | High arterial stress can induce endothelial remodeling and promote restenosis[48]. |

| | | Excessive stress may also approach the arterial wall's tensile stress limit, increasing the risk of tissue damage or fracture[49] (Figure 5B). |
|---|---|---|
| Peak stent stress | To be minimized | High stent stresses indicate the risk of mechanical failure, potentially inducing local inflammation and localized arterial damage that can cause in-stent restenosis[25,50]. |
| TAESS | 0.4 – 20 Pa | Low TAESS promotes inflammation and hyperplasia[51]; high TAESS may impair healing[52]. |
| Shear rate | < 1000 $s^{-1}$ | High shear rates might promote platelet activation and aggregation, particularly in the presence of malapposed struts, and might contribute to thrombosis[45]. |

## 3. RESULTS

### 3.1. Mechanical Performance

Overall, while both techniques revealed performance distinctions between stents, these were more marked in PSB, particularly in SB ostium clearance and arterial stress, whereas DKC tended to minimize differences between platforms.

#### 3.1.1 Strut Malapposition

Strut malapposition mostly happened close to the proximal and distal edge of the ostium (Figure 3). MV malapposition was higher for Orsiro than for Xience Sierra in both techniques (PSB: 13.0% vs 9.6%; DKC: 19.1% vs 15.5%). SB malapposition, present only in DKC, was also higher for Orsiro than Xience Sierra (10.3% vs 7.1%) (Figure 3).

#### 3.1.2 SB Ostium Clearance

SB ostium clearance was higher for Orsiro than for Xience Sierra in PSB (77% vs 64%), whereas in DKC both stents exhibited the same clearance (52%) (Figure 4). Orsiro caused greater ovalization of the SB ostium in both techniques (Ostium ellipticity – PSB: 13.4% vs 2.5%; DKC: 14.1% vs 1.2%).

#### 3.1.3 Arterial Wall Stress

Both average and peak arterial stress at the end of the procedure were greater for Orsiro compared to Xience Sierra in both techniques (Average stress - PSB: 22.8 kPa vs 20.0 kPa; DKC: 35.0 kPa vs 33.3 kPa) (Peak stress - PSB: 181.0 kPa vs 152.4 kPa; DKC: 562.9 kPa vs 533.0 kPa) (Percentage volume of arterial tissue with stress > 10 kPa – PSB: 8.4% vs 9.6%; DKC: 16.2% vs 16.0%) (Figure 5). The highest average arterial stresses were reached during KBI (Supporting Information 2).

#### 3.1.4 Peak Stent Stress

Peak MV stent stress occurred at the end of the second POT in both techniques (Figure 6). In PSB, Orsiro exhibited higher peak stress than Xience Sierra (875 MPa vs 789 MPa), corresponding to 6% and 15% below the rupture threshold of the stent material (933 MPa). In DKC, MV stent stress was similar between platforms (Orsiro vs Xience Sierra: 871 MPa vs 884 MPa), within 7% and 5% of the rupture threshold, respectively. For the SB stent in DKC, peak stress occurred at the end of the crush step and was lower overall, with Orsiro at 712 MPa (24% below threshold) and Xience Sierra at 806 MPa (14% below threshold).

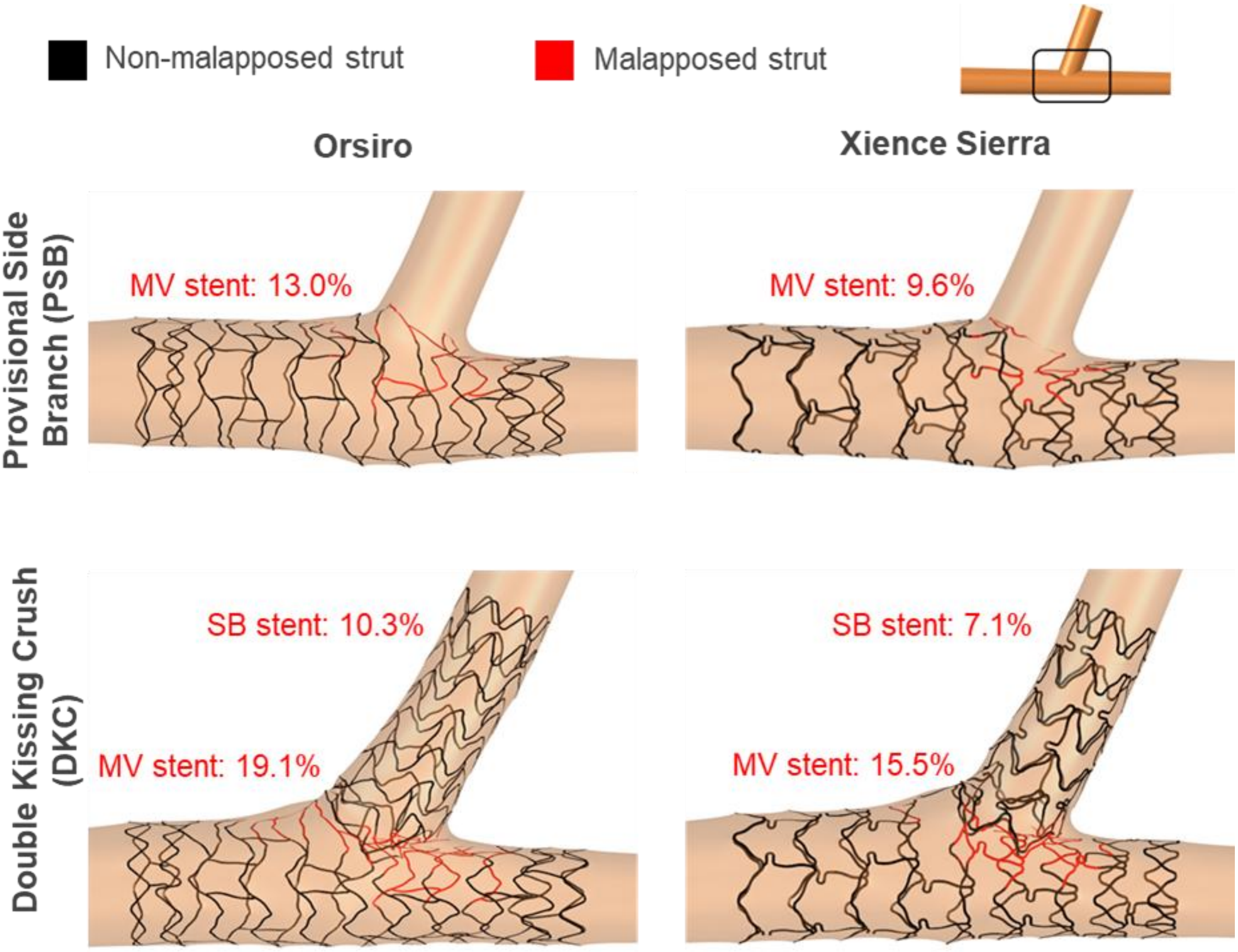

**Figure 3** – Comparison of stent malapposition in PSB (top) and DKC (bottom) techniques for Orsiro (left) and Xience Sierra (right). Malapposed struts (red) indicate regions where stent struts are not in direct contact with the arterial wall. Percentages indicate malapposition rates for each stent and branch location. Ultrathin-strut Orsiro exhibits overall a higher malapposition than thin-strut Xience Sierra, particularly in PSB. Both stents show increased malapposition in DKC, reflecting greater procedural complexity and the formation of a metallic neocarina. *DKC: Double Kissing Crush; MV: Main Vessel; PSB: Provisional Side Branch; SB: Side Branch.*

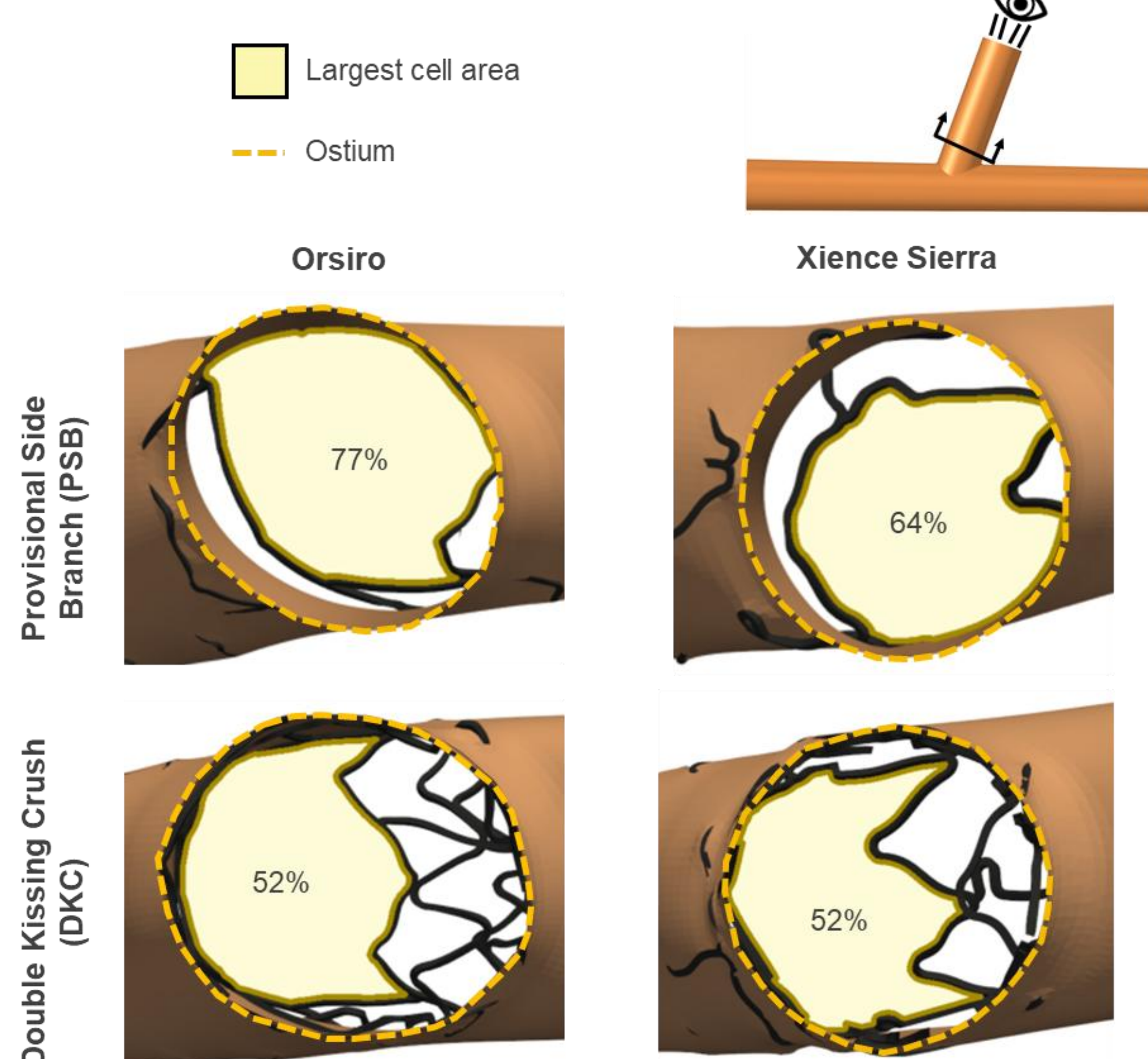

**Figure 4** – Comparison of SB ostium clearance between Orsiro (left) and Xience Sierra (right) stents deployed using PSB (top) and DKC (bottom) techniques. Cross-sectional views illustrate the largest cell area (highlighted in yellow) free of stent struts at the SB ostium, indicating regions of decreased disturbed flow. Percentage values represent the proportion of ostial opening free from metallic coverage with respect to the total ostium area. Both stents show similar ostium clearance in DKC due to the formation of the metallic neocarina, reducing ostial accessibility. Orsiro ovalized the ostium more compared to Xience Sierra. *DKC: Double Kissing Crush; PSB: Provisional Side Branch; SB: Side Branch*.

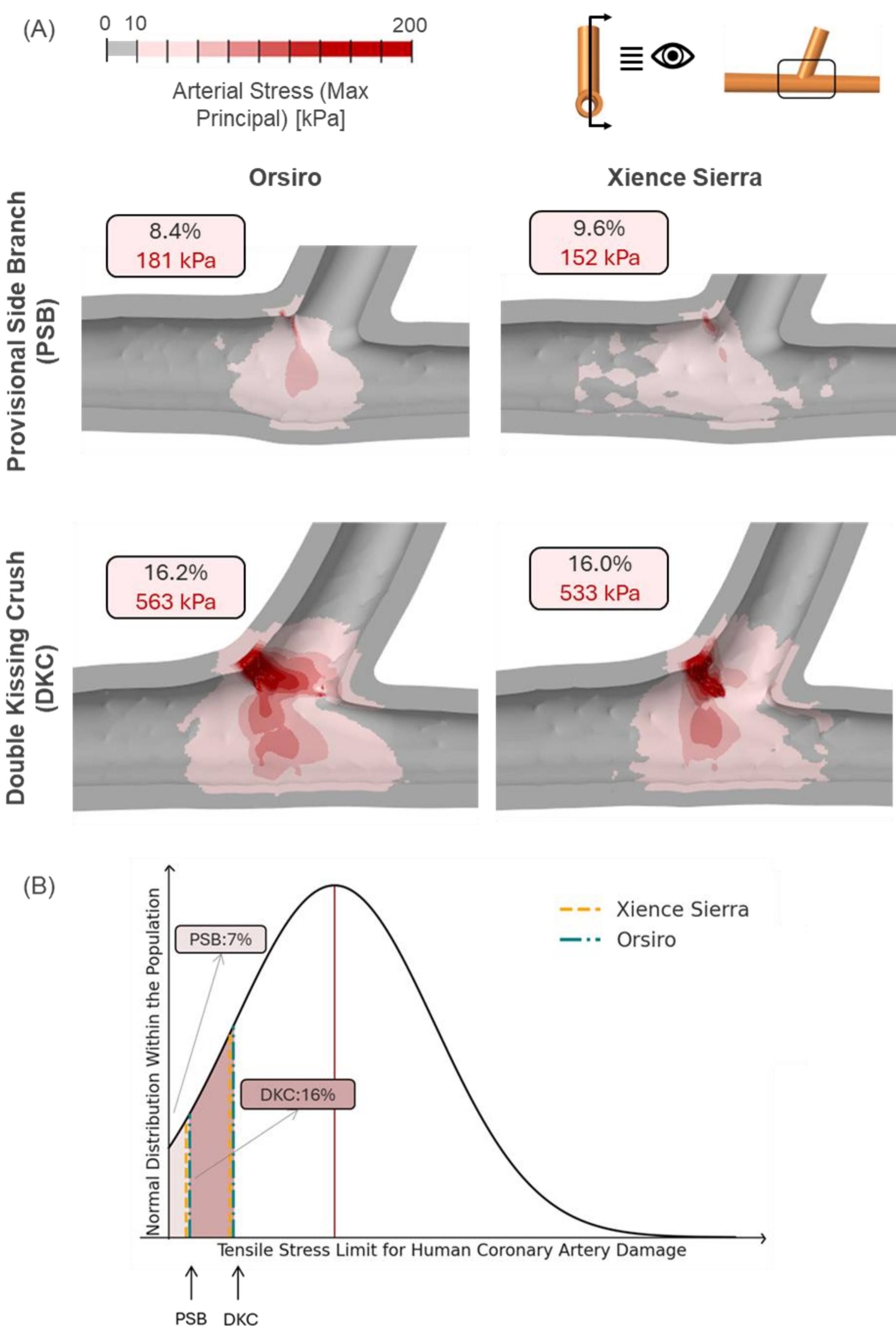

**Figure 5** – (A) Comparison of arterial wall stress (maximum principal stress) between Orsiro and Xience Sierra stents in PSB and DKC techniques. Views illustrate stress distribution at the bifurcation, with all peak stress regions concentrated at the ostium opposite the carina. For each scenario, the maximum stress value and the percentage volume of arterial wall exposed to more than 10 kPa (shadings of red) are reported. Orsiro induces higher maximum arterial stress compared to Xience Sierra at about 30 kPa, consistent across techniques. Maximum stress values increased more than 3-fold for both stents in DKC compared to PSB, while the affected vessel volume size approximately doubled. (B) Gaussian curve

illustrating how tensile stress thresholds for human coronary damage vary across the population (median 1440 kPa, standard deviation 870 kPa[49]). Vertical dashed lines mark stresses from Orsiro and Xience Sierra stents; differences between platforms are small. Stenting technique drives the risk: PSB might damage coronaries in about 7% of patients, compared with about 16% for DKC. *DKC: Double Kissing Crush; PSB: Provisional Side Branch.*

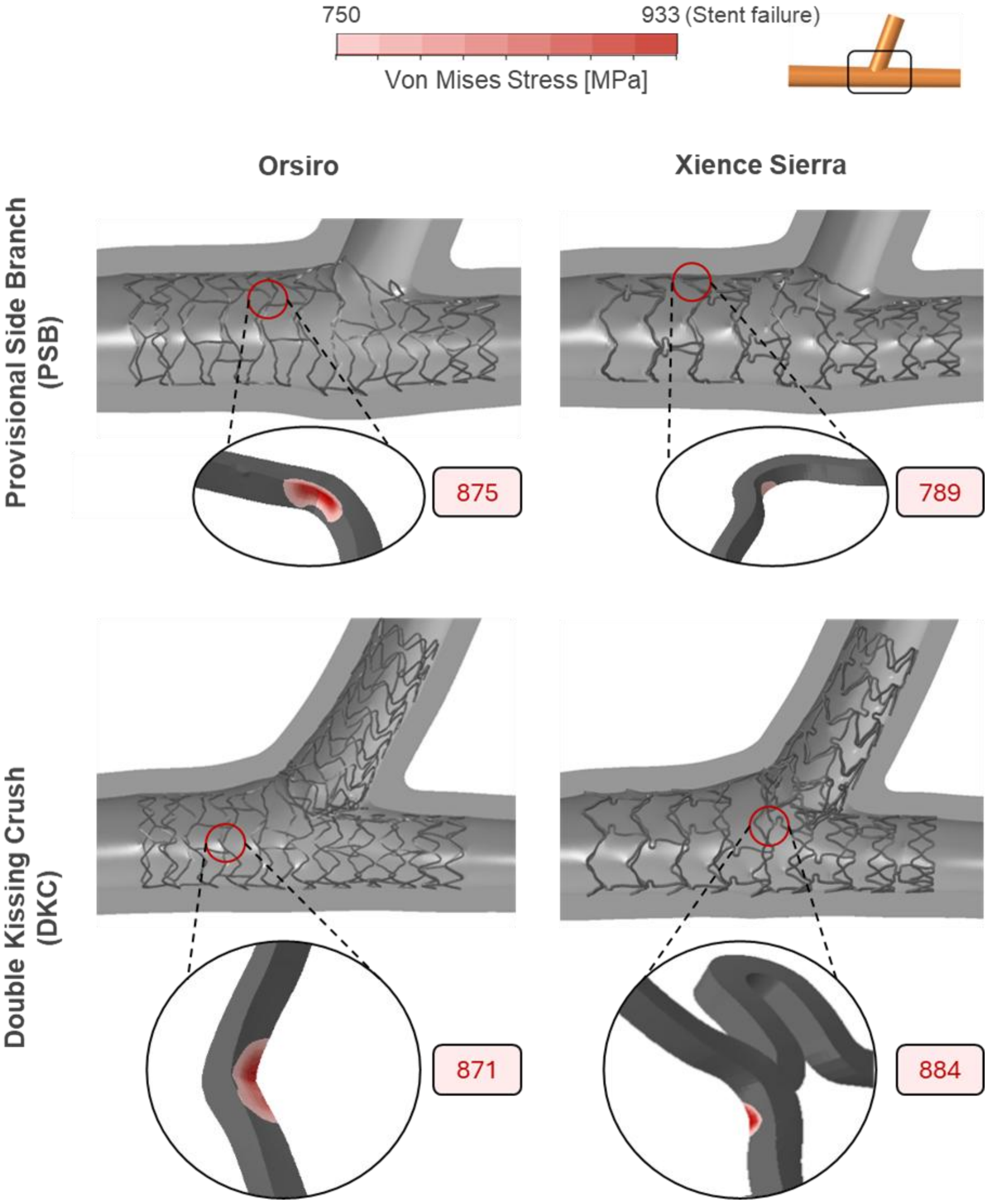

**Figure 6** – Comparison of peak MV stent stress distribution for Orsiro and Xience Sierra under PSB and DKC. Stress distributions are shown at the end of the second POT, when the maximum MV stent stress occurs. In PSB, peak stress reached 875 MPa for Orsiro and 789 MPa for Xience Sierra. In DKC, values were comparable: 871 MPa for Orsiro and 884 MPa for Xience Sierra. The stress map is scaled between 750 MPa and the material rupture threshold (933 MPa) to highlight high-stress regions. *DKC: Double Kissing Crush; MV: Main Vessel; POT: Proximal Optimization Technique; PSB: Provisional Side Branch.*

### 3.2. Hemodynamic Performance

The normalized percentage of luminal area exposed to TAESS below 0.4 Pa was lower for Orsiro compared to Xience Sierra in both techniques, with a minimal 3% and 8% difference for PSB and DKC at a 2% expected numerical error (PSB: 30.3 vs 33.6%; DKC: 28.2 vs 36.3%) (Figure 7). Across all scenarios, the maximum ESS reached 8 Pa, remaining within the physiological range. Shear rate burden indexes were 30% lower for Orsiro in PSB but 17% higher in DKC compared to Xience Sierra (PSB: 79.4 $mm^3s^{-1}$ vs 113.5 $mm^3s^{-1}$; 149.7 $mm^3s^{-1}$ vs 128.2 $mm^3s^{-1}$).

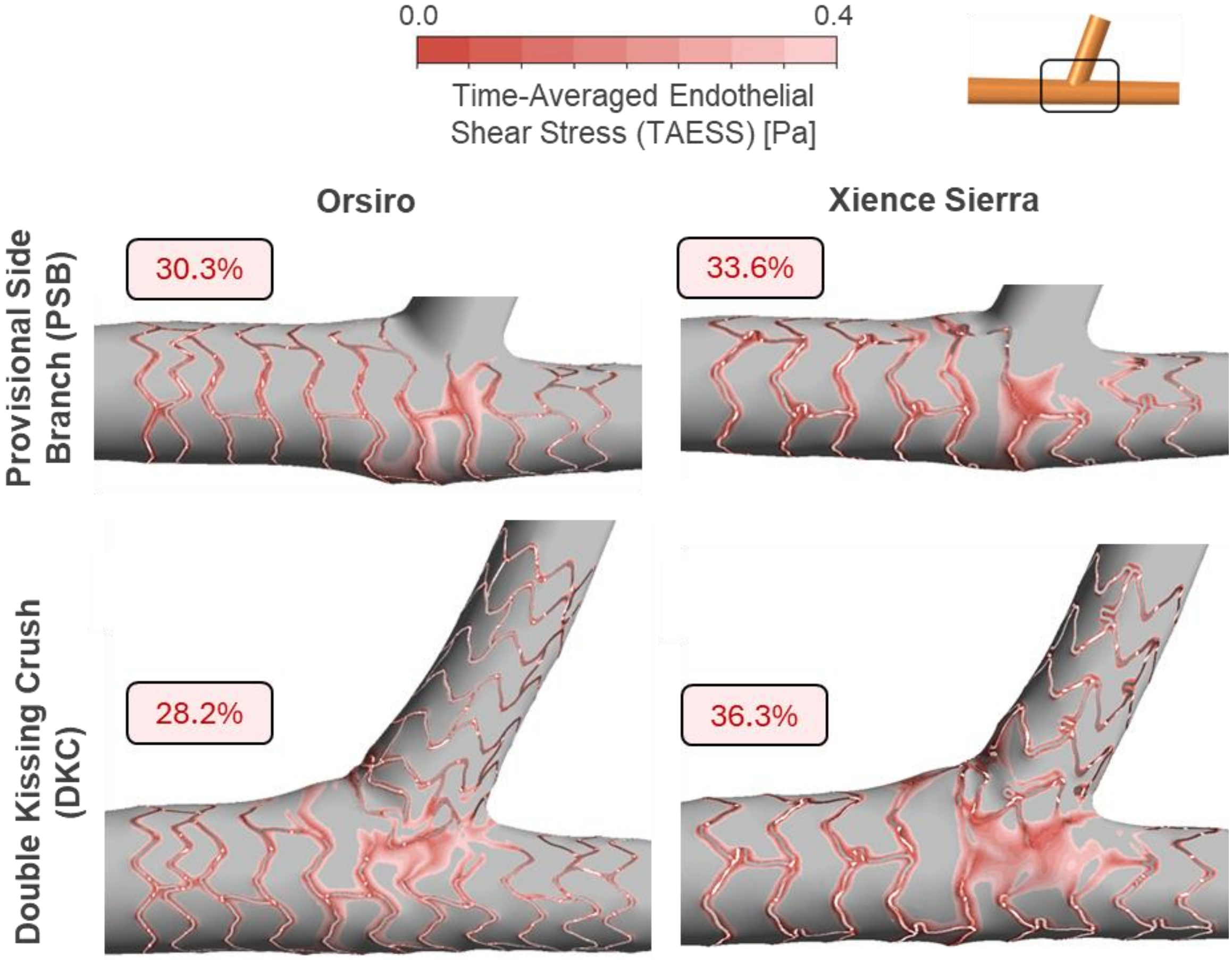

**Figure 7** – Hemodynamic comparison between Orsiro (left) and Xience Sierra (right) stents after virtual deployment using PSB (top) and DKC (bottom) techniques. Red indicates the distribution of adversely low TAESS (< 0.4 Pa) over the arterial wall. For each scenario, the percentage areas of arterial lumen exposed to adversely low TAESS are reported. Orsiro consistently demonstrates smaller low-shear-stress areas compared to Xience Sierra, indicating better hemodynamic performance, especially in DKC. *DKC: Double Kissing Crush; PSB: Provisional Side Branch; TAESS: Time-Averaged Endothelial Shear Stress.*

## 4. DISCUSSION

This study represents the first detailed mechanical and hemodynamic comparison of the widely used Orsiro and Xience Sierra stents for both the PSB and DKC bifurcation stenting techniques, incorporating clinically relevant metrics. By leveraging computational modeling, we were able to quantify key biomechanical parameters, providing new insights into the stent performance. This work presents the first comprehensive biomechanical model evaluating DKC. This novel comparative approach offers crucial insights into how stent design influences performance in complex bifurcation interventions. It also contributes to the development of a biomechanical model that supports ongoing research into this treatment strategy.

### 4.1. Effect of Stent Platform and Implant Technique

Malapposition was consistently higher with Orsiro compared to Xience Sierra, particularly in the PSB technique. This difference stems from Orsiro's thinner struts and higher radial recoil, which limit its ability to fully conform to the vessel wall. In contrast, Xience Sierra's thicker struts and ring-based design reduce recoil and offer greater radial support, promoting more stable scaffolding. Malapposition also reflects the inability of the stent to accommodate irregular or rigid arterial geometries, such as heavily calcified or fibrotic lesions. The observed malapposition rates in PSB align closely with previously reported results (3-10%)[23]. Slightly higher values in our study were expected since we tested stents with fewer connectors, which increases flexibility but can compromise wall apposition. In the DKC technique compared to PBS, malapposition increased for both stent platforms due to the presence of a second stent and additional procedural challenges, yet Xience Sierra maintained superior wall apposition overall. We can infer that the improved scaffolding and reduced recoil of Xience Sierra might prove beneficial in lesions where mechanical resistance is elevated, such as those with calcium or fibrotic plaque, where poor wall apposition could impair scaffolding and increase procedural risk.

SB ostium clearance was higher in PSB, with Orsiro showing slightly better performance due to its thinner and flexible design. These results are consistent with previous studies reporting SB ostium clearance values of 70–90%[21,32,53]. However, this advantage was accompanied by a more ovalized ostium shape, with Orsiro showing an ostium ellipticity five times higher than Xience Sierra (13.4% vs 2.5%). A less circular ostium is mechanically suboptimal, as it can lead to uneven force distribution and reduced scaffold support; in addition, it predisposes the region to disturbed flow patterns. In DKC, SB ostium clearance was comparable between the two stents, due to the formation of a metallic neocarina. SB ostium clearance values observed in DKC matched closely with findings reported by previous studies (40–52%)[22]. Similarly, ostial ovalization was more pronounced with Orsiro, showing an ostium ellipticity twelve times higher than Xience Sierra (14.1% vs 1.2%). However, it should be noted that SB ostium clearance is influenced by the landing positioning of the stent, both longitudinally and rotationally, as the alignment of the stent cell relative to the ostium determines the extent of coverage. In clinical practice, this positioning cannot be precisely controlled, and in the present simulations, stent positioning was not predetermined but resulted naturally from the deployment process. This challenge is particularly relevant for Orsiro, which has lower radiopacity due to its thinner struts, making precise positioning under fluoroscopy more difficult[54].

Arterial wall stress peaked during KBI, occurring at the ostium opposite the carina, where Orsiro induced higher stress levels. Notably, at the end of the PSB procedure, arterial wall stress remained elevated for Orsiro compared to Xience Sierra. This is likely due to Orsiro's helical structure, which interacts differently with the balloons and bifurcation, and higher ovalization of the ostium, altering the distribution of stress across the arterial wall. Although thinner struts are generally expected to reduce arterial wall stress[55,56], Orsiro's specific helical configuration appeared to elevate arterial wall stress overall. While a precise threshold for adverse arterial stress limit has not been established to date, elevated arterial wall stress has been implicated in intimal hyperplasia and delayed healing[48]. In this context, the comparative differences in arterial wall stress suggest a greater potential for adverse remodeling for Orsiro, especially in DKC. However, simulated arterial stresses remained well below the mean stress limit before rupture reported after tensile tests of 14 *ex vivo* human coronary arteries (mean failure stress 1440 kPa) (Figure 5B)[49]. Still, our simulations suggest that approximately 7% of patients undergoing PSB, and 16% undergoing DKC could experience tissue injury. Nonetheless, both techniques appear safe for the majority of patients, as the resulting stresses remain below the damage threshold for most of the population. The main determinant of tissue injury risk was the stenting technique, not the stent platform, since differences in the maximum arterial stress between Orsiro and Xience Sierra were smaller.

Stent stresses in the MV were high for both platforms, with Orsiro generally exhibiting slightly greater peak values. In all cases, peak stress occurred at the ostium during the second POT, reaching values within 5-7% of the material's rupture threshold. While no fracture was observed in the simulation, this narrow margin indicates that the stent is already operating near its mechanical limit. If overexpansion is applied during POT to reduce malapposition, it may exceed this limit, increasing the risk of strut fracture. Although the crush step is often considered mechanically demanding, our findings show that POT imposes greater stress on the stent. Since POT is routinely performed to address MV-to-DV diameter mismatch, the mechanical burden it introduces cannot be avoided and must be carefully managed. These results highlight the importance of selecting a stent that is appropriately sized for the target vessel and emphasize the need for precise procedural planning to minimize mechanical risk. Overall, Xience Sierra design resulted in improved wall apposition and lower arterial stresses, and better circularity of the SB ostium, making it a suitable choice when mechanical performance is a priority, such as in scenarios requiring high radial strength and scaffolding properties. This aligns with clinical studies reporting Xience Sierra's superior performance in challenging anatomical conditions involving heavily calcified lesions or significant vessel diameter mismatches[10,57]. However, the differences between the two stent platforms drastically reduced in the DKC setting.

The overall hemodynamic performance of the Orsiro stent was more favorable, due to its thinner struts and helical configuration, which promoted smoother flow compared to the slightly thicker-strutted, ring-based design of Xience Sierra. This advantage was maintained in both PSB and DKC despite Orsiro's longer length. Specifically, Orsiro significantly reduced the luminal areas exposed to adversely low TAESS (< 0.4 Pa) by generating higher overall mean TAESS in the stented region. This is linked to reduced inflammation and neointimal hyperplasia risk[51,52]. This improvement is consistent even before adjusting for Orsiro's slightly longer stent length (13 mm vs 12 mm for Xience) (Supporting Information 2). Orsiro's design may therefore mitigate restenosis risk in patients with predisposition, such as those with diabetes, long lesions, or small vessels, in alignment with clinical reports[5-8,55]. In contrast, high shear rates, which are associated with platelet activation and thrombosis[45], showed a more technique-dependent pattern. In PSB, Orsiro exhibited a lower thrombogenic shear burden than Xience Sierra, consistent with its better ostial scaffolding and despite slightly greater malapposition. In DKC, however, Orsiro showed higher shear rate burden, likely due to its increased malapposition and reduced SB ostium clearance. These results suggest that Orsiro may reduce both restenotic and thrombotic risk in single-stent procedures, but its advantage is diminished in two-stent configurations, where mechanical factors such as malapposition play a more dominant role.

### 4.2. Implications for Clinical Practice

The choice of stenting technique significantly affects procedural outcomes in PCI and should be prioritized over stent platform selection. The PSB technique led to better outcomes compared to DKC across all metrics, including lower arterial and stent stresses, malapposition, and better SB ostium clearance. PSB revealed performance differences between stent platforms, making device selection more influential compared to DKC. Xience Sierra's reduced radial recoil and improved strut apposition may be advantageous in cases with calcification, wide bifurcation angles (prevalent in males[36,58]), or large bifurcation diameter mismatch. Orsiro, by contrast, demonstrated better hemodynamic performance, potentially reducing adverse flow patterns, especially relevant for patients with long lesions, small vessels (prevalent in females[36,58]), or diabetes. These findings highlight that in PSB, lesion-specific stent selection plays a critical role in optimizing outcomes.

The DKC technique imposed a greater mechanical burden due to its procedural complexity and the use of two overlapping stents compared to PBS. This led to worse overall outcomes for both platforms, such as higher arterial and stent stresses, greater malapposition, and reduced ostial clearance. In fact, the DKC strategy mitigated the mechanical performance differences between stents. Still, Orsiro retained a hemodynamic advantage, which may make it the preferred option for DKC. This possibility is currently being investigated in an ongoing clinical trial comparing these stents in true bifurcation lesions treated with DKC (NCT05200637).

Overall, neither stent is superior across all performance metrics, reinforcing findings from clinical trials that report no definitive stent preference in general patient populations[2-4]. Instead, our results support a tailored approach: in PSB, device choice should align with lesion characteristics; in DKC, Orsiro's hemodynamic benefits may be advantageous, although the technique itself remains the dominant determinant of outcome. This underscores the need to consider both anatomy and procedural strategy when selecting a stent, since they mutually influence each other.

Finally, optimizing critical procedural steps, particularly POT and KBI, is essential to minimize arterial and stent stresses and reduce the incidence of malapposition. Our findings highlight that suboptimal performance in these steps can worsen the clinical outcome. In fact, correct execution of KBI minimizes arterial stress, while properly conducted POT addresses MV-to-DV diameter mismatches and mitigates risks associated with stent overexpansion.

### 4.3. Limitations

This combined FEA and CFD study employed necessary simplifying assumptions, including the exclusion of plaque and the use of generalized vessel geometry. However, these choices reflect common practice in biomechanical analyses, driven by the uncertainty of true plaque component material properties, and large variability of plaque heterogeneity[17]. Such standardization enhances robustness and enables direct comparison across stent designs and techniques. Moreover, to achieve greater representation of a population, the present generalized bifurcation anatomy is preferable. The stent models, derived from µCT imaging, are geometrically simplified, which may influence local stress-strain predictions and mechanical outcomes such as radial stiffness or recoil. Due to a lack of published data on the specific alloys used in these clinical stents, the device material behavior was modeled using available literature data for alloys of similar composition. While this may not fully capture the processed mechanical response of the deployed devices, the approach ensures consistency across simulations. Future studies should examine the impact of variations in stent positioning and orientation to evaluate the translational relevance of numerical approaches further. Integrating plaque characteristics and patient-specific vessel geometries may also generate additional, nuanced population group-specific insights and would be a significant computational undertaking.

## 5. CONCLUSION

This study is the first to comprehensively evaluate the mechanical and hemodynamic performances of Orsiro and Xience Sierra stents for both PSB and DKC techniques. In PSB, the influence of stent platforms on key performance metrics is evident, whereby Xience Sierra's reduced radial recoil and improved strut apposition may benefit cases with calcification, wide bifurcation angles, or large diameter mismatch. Orsiro's favorable hemodynamics, by contrast, may be more appropriate for patients with small vessels, long lesions, or elevated risk of restenosis. On the other hand, under the more complex DKC technique, performance differences between stents became less pronounced, as the procedural burden tended to reduce structural outcomes. However, Orsiro retained a hemodynamic advantage, possibly making it preferable for DKC. These findings support a tailored approach to stent choice: device selection plays a larger role in PSB, while outcomes in DKC are driven more by the technique itself.

Finally, the findings emphasize the critical role of procedural technique, showcasing that a precise execution of steps such as KBI and POT is essential for minimizing arterial and stent stresses. Notably, POT emerged as more influential than the crush step in determining stent integrity, due to its impact on overexpansion and MV-to-DV diameter mismatch. These insights underscore the need for careful technique selection and meticulous execution to fully leverage each stent's unique design advantages.

## 6. ACKNOWLEDGEMENTS

This research includes computations performed on the computational cluster Katana (supported by Research Technology Services at UNSW Sydney) and from the computational cluster Gadi (supported by National Computational Infrastructure, an NCRIS-enabled facility supported by the Australian Government).

## 7. DISCLOSURE

During the preparation of this work the authors used ChatGPT 4.0 and Grammarly to detect grammatical errors. After using this tool/service, the author(s) reviewed and edited the content as needed and take(s) full responsibility for the content of the publication.

## 8. REFERENCES

1. Goel R, Spirito A, Gao M, Vogel B, N Kalkman D, Mehran R. Second-generation everolimus-eluting intracoronary stents: a comprehensive review of the clinical evidence. *Future Cardiology*. 2024;20:103-116. doi: 10.2217/fca-2023-0092
2. Yamaji K, Zanchin T, Zanchin C, Stortecky S, Koskinas KC, Hunziker L, Praz F, Blöchlinger S, Moro C, Moschovitis A, et al. Unselected Use of Ultrathin Strut Biodegradable Polymer Sirolimus-Eluting Stent Versus Durable Polymer Everolimus-Eluting Stent for Coronary Revascularization. *Circulation: Cardiovascular Interventions*. 2018;11:e006741. doi: doi:10.1161/CIRCINTERVENTIONS.118.006741

3. Nakamura M, Kadota K, Nakagawa Y, Tanabe K, Ito Y, Amano T, Maekawa Y, Takahashi A, Shiode N, Otsuka Y, et al. Ultrathin, Biodegradable-Polymer Sirolimus-Eluting Stent vs Thin, Durable-Polymer Everolimus-Eluting Stent. *JACC: Cardiovascular Interventions*. 2022;15:1324-1334. doi: https://doi.org/10.1016/j.jcin.2022.05.028
4. Lefèvre T, Haude M, Neumann F-J, Stangl K, Skurk C, Slagboom T, Sabaté M, Goicolea J, Barragan P, Cook S, et al. Comparison of a Novel Biodegradable Polymer Sirolimus-Eluting Stent With a Durable Polymer Everolimus-Eluting Stent. *JACC: Cardiovascular Interventions*. 2018;11:995-1002. doi: doi:10.1016/j.jcin.2018.04.014
5. Pilgrim T, Piccolo R, Heg D, Roffi M, Tüller D, Muller O, Moarof I, Siontis GCM, Cook S, Weilenmann D, et al. Ultrathin-strut, biodegradable-polymer, sirolimus-eluting stents versus thin-strut, durable-polymer, everolimus-eluting stents for percutaneous coronary revascularisation: 5-year outcomes of the BIOSCIENCE randomised trial. *The Lancet*. 2018;392:737-746. doi: https://doi.org/10.1016/S0140-6736(18)31715-X
6. Pilgrim T, Muller O, Heg D, Roffi M, Kurz DJ, Moarof I, Weilenmann D, Kaiser C, Tapponnier M, Losdat S, et al. Biodegradable- Versus Durable-Polymer Drug-Eluting Stents for STEMI: Final 2-Year Outcomes of the BIOSTEMI Trial. *JACC: Cardiovascular Interventions*. 2021;14:639-648. doi: https://doi.org/10.1016/j.jcin.2020.12.011
7. Buiten RA, Ploumen EH, Zocca P, Doggen CJM, van der Heijden LC, Kok MM, Danse PW, Schotborgh CE, Scholte M, de Man FHAF, et al. Outcomes in Patients Treated With Thin-Strut, Very Thin-Strut, or Ultrathin-Strut Drug-Eluting Stents in Small Coronary Vessels: A Prespecified Analysis of the Randomized BIO-RESORT Trial. *JAMA Cardiology*. 2019;4:659-669. doi: 10.1001/jamacardio.2019.1776
8. Forrestal BJ, Case BC, Yerasi C, Garcia-Garcia HM, Waksman R. The Orsiro Ultrathin, Bioresorbable-Polymer Sirolimus-Eluting Stent: A Review of Current Evidence. *Cardiovascular Revascularization Medicine*. 2020;21:540-548. doi: https://doi.org/10.1016/j.carrev.2019.12.039
9. Sherbet DP, Christopoulos G, Karatasakis A, Danek BA, Kotsia A, Navara R, Michael TT, Roesle M, Rangan BV, Haagen D, et al. Optical coherence tomography findings after chronic total occlusion interventions: Insights from the “AngiographiC evaluation of the everolimus-eluting stent in chronic Total occlusions” (ACE-CTO) study (NCT01012869). *Cardiovascular Revascularization Medicine*. 2016;17:444-449. doi: https://doi.org/10.1016/j.carrev.2016.04.002
10. Zivelonghi C, Teeuwen K, Agostoni P, van der Schaaf RJ, Ribichini F, Adriaenssens T, Kelder JC, Tijssen JGP, Henriques JPS, Suttorp MJ. Impact of ultra-thin struts on restenosis after chronic total occlusion recanalization: Insights from the randomized PRISON IV trial. *Journal of Interventional Cardiology*. 2018;31:580-587. doi: https://doi.org/10.1111/joic.12516
11. Zivelonghi C, Agostoni P, Teeuwen K, van der Schaaf RJ, Henriques JPS, Vermeersch P, Bosschaert MAR, Kelder JC, Tijssen JGP, Suttorp MJ. 3-Year Clinical Outcomes of the PRISON-IV Trial: Ultrathin Struts Versus Conventional Drug-Eluting Stents in Total Coronary Occlusions. *JACC Cardiovasc Interv*. 2019;12:1747-1749. doi: 10.1016/j.jcin.2019.05.044
12. Kapoor A, Jepson N, Bressloff NW, Loh P, Ray T, Beier S. The road to the ideal stent: A review of stent design optimisation methods, findings, and opportunities. *Materials & Design*. 2023;237:112556. doi: https://doi.org/10.1016/j.matdes.2023.112556
13. Gharleghi R, Wright H, Luvio V, Jepson N, Luo Z, Senthurnathan A, Babaei B, Prusty BG, Ray T, Beier S. A multi-objective optimization of stent geometries. *Journal of Biomechanics*. 2021;125:110575. doi: https://doi.org/10.1016/j.jbiomech.2021.110575
14. Watson T, Webster MWI, Ormiston JA, Ruygrok PN, Stewart JT. Long and short of optimal stent design. *Open Heart*. 2017;4:e000680. doi: 10.1136/openhrt-2017-000680

15. Kim DB, Choi H, Joo SM, Kim HK, Shin JH, Hwang MH, Choi J, Kim D-G, Lee KH, Lim CH, et al. A Comparative Reliability and Performance Study of Different Stent Designs in Terms of Mechanical Properties: Foreshortening, Recoil, Radial Force, and Flexibility. *Artificial Organs*. 2013;37:368-379. doi: https://doi.org/10.1111/aor.12001
16. Beier S, Ormiston J, Webster M, Cater J, Norris S, Medrano-Gracia P, Young A, Cowan B. Hemodynamics in Idealized Stented Coronary Arteries: Important Stent Design Considerations. *Annals of Biomedical Engineering*. 2016;44:315-329. doi: 10.1007/s10439-015-1387-3
17. Colombo A, Chiastra C, Gallo D, Loh PH, Dokos S, Zhang M, Keramati H, Carbonaro D, Migliavacca F, Ray T, et al. Advancements in Coronary Bifurcation Stenting Techniques: Insights From Computational and Bench Testing Studies. *International Journal for Numerical Methods in Biomedical Engineering*. 2025;41:e70000. doi: https://doi.org/10.1002/cnm.70000
18. Albiero R, Burzotta F, Lassen JF, Lefèvre T, Banning AP, Chatzizisis YS, Johnson TW, Ferenc M, Pan M, Darremont O. Treatment of coronary bifurcation lesions, part I: implanting the first stent in the provisional pathway. The 16th expert consensus document of the European Bifurcation Club. *EuroIntervention*. 2022;18:e362. doi: 10.4244/EIJ-D-22-00165
19. Chen S-L, Sheiban I, Xu B, Jepson N, Paiboon C, Zhang J-J, Ye F, Sansoto T, Kwan TW, Lee M, et al. Impact of the Complexity of Bifurcation Lesions Treated With Drug-Eluting Stents: The DEFINITION Study (Definitions and impact of complEx biFurcation lesIons on clinical outcomes after percutaNeous coronary IntervenTIOn using drug-eluting steNts). *JACC: Cardiovascular Interventions*. 2014;7:1266-1276. doi: https://doi.org/10.1016/j.jcin.2014.04.026
20. Lassen JF, Albiero R, Johnson TW, Burzotta F, Lefevre T, Iles TL, Pan M, Banning AP, Chatzizisis YS, Ferenc M, et al. Treatment of coronary bifurcation lesions, part II: implanting two stents. The 16th expert consensus document of the European Bifurcation Club. *EuroIntervention*. 2022. doi: 10.4244/EIJ-D-22-00166
21. Rigatelli G, Zuin M, Chiastra C, Burzotta F. Biomechanical Evaluation of Different Balloon Positions for Proximal Optimization Technique in Left Main Bifurcation Stenting. *Cardiovasc Revasc Med*. 2020;21:1533-1538. doi: 10.1016/j.carrev.2020.05.028
22. Paradies V, Ng J, Lu S, Bulluck H, Burzotta F, Chieffo A, Ferenc M, Wong PE, Hausenloy DJ, Foin N, et al. T and Small Protrusion (TAP) vs Double-Kissing Crush Technique: Insights From In Vitro Models. *Cardiovascular Revascularization Medicine*. 2021;24:11-17. doi: 10.1016/j.carrev.2020.09.013
23. Mortier P, Hikichi Y, Foin N, De Santis G, Segers P, Verhegghe B, De Beule M. Provisional Stenting of Coronary Bifurcations: Insights Into Final Kissing Balloon Post-Dilation and Stent Design by Computational Modeling. *JACC: Cardiovascular Interventions*. 2014;7:325-333. doi: https://doi.org/10.1016/j.jcin.2013.09.012
24. Burzotta F, Louvard Y, Lassen JF, Lefèvre T, Finet G, Collet C, Legutko J, Lesiak M, Hikichi Y, Albiero R, et al. Percutaneous coronary intervention for bifurcation coronary lesions using optimised angiographic guidance: the 18th consensus document from the European Bifurcation Club. *EuroIntervention*. 2024;20:e915-e926. doi: 10.4244/eij-d-24-00160
25. Gasior P, Lu S, Ng CKJ, Toong WYD, Wong EHP, Foin N, Kedhi E, Wojakowski W, Ang HY. Comparison of overexpansion capabilities and thrombogenicity at the side branch ostia after implantation of four different drug eluting stents. *Scientific Reports*. 2020;10:20791. doi: 10.1038/s41598-020-75836-6
26. Morlacchi S, Chiastra C, Cutri E, Zunino P, Burzotta F, Formaggia L, Dubini G, Migliavacca F. Stent deformation, physical stress, and drug elution obtained with provisional stenting, conventional culotte and Tryton-based culotte to treat bifurcations:

a virtual simulation study. *EuroIntervention*. 2014;9:1441-1453. doi: 10.4244/EIJV9I12A242
27. Zhao S, Wu W, Samant S, Khan B, Kassab GS, Watanabe Y, Murasato Y, Sharzehee M, Makadia J, Zolty D. Patient-specific computational simulation of coronary artery bifurcation stenting. *Scientific reports*. 2021;11:1-17. doi: 10.1038/s41598-021-95026-2
28. Raben JS, Morlacchi S, Burzotta F, Migliavacca F, Vlachos PP. Local Blood Flow Patterns in Stented Coronary Bifurcations: An Experimental and Numerical Study. *Journal of Applied Biomaterials & Functional Materials*. 2015;13:116-126. doi: 10.5301/jabfm.5000217
29. Iannaccone F, Chiastra C, Antonios K, Francesco M, Frank JHG, Patrick S, Peter M, Benedict V, Gabriele D, Matthieu De B, et al. Impact of plaque type and side branch geometry on side branch compromise after provisional stent implantation: a simulation study. *EuroIntervention*. 2017;13:e236-e245. doi: 10.4244/EIJ-D-16-00498
30. Burzotta F, Mortier P, Trani C. Characteristics of drug-eluting stent platforms potentially influencing bifurcated lesion provisional stenting procedure. *Eurointervention: Journal of Europcr in Collaboration with the Working Group on Interventional Cardiology of the European Society of Cardiology*. 2014;10:124-132. doi: 10.4244/EIJV10I1A19
31. Morlacchi S, Chiastra C, Gastaldi D, Pennati G, Dubini G, Migliavacca F. Sequential Structural and Fluid Dynamic Numerical Simulations of a Stented Bifurcated Coronary Artery. *Journal of Biomechanical Engineering*. 2011;133. doi: 10.1115/1.4005476
32. Foin N, Torii R, Mortier P, De Beule M, Viceconte N, Chan PH, Davies JE, Xu XY, Krams R, Di Mario C. Kissing Balloon or Sequential Dilation of the Side Branch and Main Vessel for Provisional Stenting of Bifurcations: Lessons From Micro-Computed Tomography and Computational Simulations. *JACC: Cardiovascular Interventions*. 2012;5:47-56. doi: https://doi.org/10.1016/j.jcin.2011.08.019
33. Gastaldi D, Morlacchi S, Nichetti R, Capelli C, Dubini G, Petrini L, Migliavacca F. Modelling of the provisional side-branch stenting approach for the treatment of atherosclerotic coronary bifurcations: effects of stent positioning. *Biomechanics and Modeling in Mechanobiology*. 2010;9:551-561. doi: 10.1007/s10237-010-0196-8
34. Chiastra C, Grundeken MJ, Collet C, Wu W, Wykrzykowska JJ, Pennati G, Dubini G, Migliavacca F. Biomechanical Impact of Wrong Positioning of a Dedicated Stent for Coronary Bifurcations: A Virtual Bench Testing Study. *Cardiovasc Eng Technol*. 2018;9:415-426. doi: 10.1007/s13239-018-0359-9
35. Grundeken MJ, Chiastra C, Wu W, Wykrzykowska JJ, De Winter RJ, Dubini G, Migliavacca F. Differences in rotational positioning and subsequent distal main branch rewiring of the Tryton stent: An optical coherence tomography and computational study. *Catheterization and Cardiovascular Interventions*. 2018;92:897-906. doi: https://doi.org/10.1002/ccd.27567
36. Medrano-Gracia P, Ormiston J, Webster M, Beier S, Young A, Ellis C, Wang C, Smedby Ö, Cowan B. A computational atlas of normal coronary artery anatomy. *EuroIntervention: journal of EuroPCR in collaboration with the Working Group on Interventional Cardiology of the European Society of Cardiology*. 2016;12:845-854. doi: 10.4244/EIJV12I7A139
37. Kapoor A, Ray T, Jepson N, Beier S. Comprehensive Geometric Parameterization and Computationally Efficient 3D Shape Matching Optimization of Realistic Stents. *Journal of Mechanical Design*. 2024;147. doi: 10.1115/1.4066961
38. Kapoor A, Ray T, Jepson N, Beier S. A Surrogate-Assisted Multiconcept Optimization Framework for Real-World Engineering Design. *Journal of Mechanical Design*. 2025;147. doi: 10.1115/1.4068404
39. Kapoor A, Ray, T., Jepson, N., & Beier, S. 3D geometry builder for Independent Ring and Helical Stent designs. In: Journal of Mechanical Design; 2025.

40. Burzotta F, Lassen JF, Louvard Y, Lefèvre T, Banning AP, Daremont O, Pan M, Hildick-Smith D, Chieffo A, Chatzizisis YS, et al. European Bifurcation Club white paper on stenting techniques for patients with bifurcated coronary artery lesions. *Catheterization and Cardiovascular Interventions*. 2020;96:1067-1079. doi: https://doi.org/10.1002/ccd.29071
41. Razavi A, Shirani E, Sadeghi MR. Numerical simulation of blood pulsatile flow in a stenosed carotid artery using different rheological models. *Journal of Biomechanics*. 2011;44:2021-2030. doi: https://doi.org/10.1016/j.jbiomech.2011.04.023
42. Shen C, Gharleghi R, Li DD, Stevens M, Dokos S, Beier S. Secondary flow in bifurcations - Important effects of curvature, bifurcation angle and stents. *J Biomech*. 2021;129:110755. doi: 10.1016/j.jbiomech.2021.110755
43. Beier S, Ormiston J, Webster M, Cater J, Norris S, Medrano-Gracia P, Young A, Cowan B. Impact of bifurcation angle and other anatomical characteristics on blood flow–A computational study of non-stented and stented coronary arteries. *Journal of biomechanics*. 2016;49:1570-1582. doi: 10.1016/j.jbiomech.2016.03.038
44. van der Giessen AG, Groen HC, Doriot P-A, de Feyter PJ, van der Steen AFW, van de Vosse FN, Wentzel JJ, Gijsen FJH. The influence of boundary conditions on wall shear stress distribution in patients specific coronary trees. *Journal of Biomechanics*. 2011;44:1089-1095. doi: https://doi.org/10.1016/j.jbiomech.2011.01.036
45. Foin N, Gutiérrez-Chico JL, Nakatani S, Torii R, Bourantas CV, Sen S, Nijjer S, Petraco R, Kousera C, Ghione M, et al. Incomplete Stent Apposition Causes High Shear Flow Disturbances and Delay in Neointimal Coverage as a Function of Strut to Wall Detachment Distance. *Circulation: Cardiovascular Interventions*. 2014;7:180-189. doi: doi:10.1161/CIRCINTERVENTIONS.113.000931
46. Foin N, Lu S, Ng J, Bulluck H, Hausenloy DJ, Wong PE, Virmani R, Joner M. Stent malapposition and the risk of stent thrombosis: mechanistic insights from an in vitro model. *EuroIntervention*. 2017;13:e1096-e1098. doi: 10.4244/eij-d-17-00381
47. Paradies V, Lu S, Ng J, Ang HY, Joner M, Foin N. Thrombogenicity at the jailed side branch ostia in the provisional stenting technique: insights from an in vitro model. *EuroIntervention*. 2018;14:826-827. doi: 10.4244/eij-d-18-00003
48. Birukov KG. Cyclic stretch, reactive oxygen species, and vascular remodeling. *Antioxid Redox Signal*. 2009;11:1651-1667. doi: 10.1089/ars.2008.2390
49. Karimi A, Navidbakhsh M, Shojaei A, Faghihi S. Measurement of the uniaxial mechanical properties of healthy and atherosclerotic human coronary arteries. *Materials Science and Engineering: C*. 2013;33:2550-2554.
50. Shlofmitz E, Iantorno M, Waksman R. Restenosis of Drug-Eluting Stents. *Circulation: Cardiovascular Interventions*. 2019;12:e007023. doi: doi:10.1161/CIRCINTERVENTIONS.118.007023
51. Malek AM, Alper SL, Izumo S. Hemodynamic Shear Stress and Its Role in Atherosclerosis. *JAMA*. 1999;282:2035-2042. doi: 10.1001/jama.282.21.2035
52. Dolan JM, Kolega J, Meng H. High Wall Shear Stress and Spatial Gradients in Vascular Pathology: A Review. *Annals of Biomedical Engineering*. 2013;41:1411-1427. doi: 10.1007/s10439-012-0695-0
53. Foin N, Torii R, Alegria E, Sen S, Petraco R, Nijjer S, Ghione M, Davies JE, Di Mario C. Location of side branch access critically affects results in bifurcation stenting: Insights from bench modeling and computational flow simulation. *International Journal of Cardiology*. 2013;168:3623-3628. doi: https://doi.org/10.1016/j.ijcard.2013.05.036
54. Piriou P-G, Guérin P, Bonin M, Plessis J, Letocart V, Manigold T, Guy B, Jordana F. Radiopacity of Coronary Stents, an In Vitro Comparative Study. *Cardiovascular Engineering and Technology*. 2020;11:719-724. doi: 10.1007/s13239-020-00492-w

55. Pilgrim T, Rothenbühler M, Siontis GCM, Kandzari DE, Iglesias JF, Asami M, Lefèvre T, Piccolo R, Koolen J, Saito S, et al. Biodegradable polymer sirolimus-eluting stents vs durable polymer everolimus-eluting stents in patients undergoing percutaneous coronary intervention: A meta-analysis of individual patient data from 5 randomized trials. *American Heart Journal*. 2021;235:140-148. doi: https://doi.org/10.1016/j.ahj.2021.02.009
56. Wiesent L, Spear A, Nonn A. Computational analysis of the effects of geometric irregularities on the interaction of an additively manufactured 316L stainless steel stent and a coronary artery. *Journal of the Mechanical Behavior of Biomedical Materials*. 2022;125:104878. doi: https://doi.org/10.1016/j.jmbbm.2021.104878
57. Buiten RA, Ploumen EH, Zocca P, Doggen CJM, van Houwelingen KG, Danse PW, Schotborgh CE, Stoel MG, Scholte M, Linssen GCM, et al. Three contemporary thin-strut drug-eluting stents implanted in severely calcified coronary lesions of participants in a randomized all-comers trial. *Catheterization and Cardiovascular Interventions*. 2020;96:E508-E515. doi: https://doi.org/10.1002/ccd.28886
58. Medrano-Gracia P, Ormiston JA, Webster MW, Beier S, Ellis C, Wang C, Smedby Ö, Young AA, Cowan BR. A Study of Coronary Bifurcation Shape in a Normal Population. *Journal of Cardiovascular Translational Research*. 2016;10:82 - 90.

**Supporting Information 1**

For the finite element model, stent deployment was simulated using FEA with Abaqus Explicit (Dassault Systèmes, Providence, USA). The arterial wall was divided into three layers, namely intima, media, and adventitia, each modeled as isotropic hyperelastic materials using a sixth-order reduced polynomial formulation with a density of 1120 $kg/m^3$ [1]. The arterial wall was meshed using C3D8R hexahedral elements. The stent material was a cobalt-chromium alloy, which was modeled with isotropic elastic-plastic behavior with a density of 8000 $kg/m^3$, also meshed with C3D8R elements[2]. Non-compliant balloons were modelled upon the geometry of the Accuforce catheter (Terumo, Japan) as membrane structures using M3D4R elements, with a density of 1000 $kg/m^3$ and Poisson's ratio of 0.45. The Young's modulus of each balloon was calibrated against manufacturer-provided pressure-diameter curves to replicate inflation behavior. The complete material property parameters used in the model are summarized in Table S1. Boundary conditions included fixing the vessel ends to constrain arterial movement and applying incremental pressure loading to simulate realistic balloon expansion for both the PSB and DKC techniques. Crimping onto a delivery balloon was simulated prior to deployment to replicate the mechanical preconditioning of the stent. Balloon rewiring positioning was achieved using the HyperMorph tool in HyperMesh (Altair Engineering, Troy, MI, USA). All simulations were conducted under quasi-static conditions using mass scaling, and the kinetic energy-to-internal energy ratio was monitored to remain below 5%, ensuring valid quasi-static assumptions.

Table S2 – Material coefficients used in the finite element simulations for arterial layers, stent, and balloons[1].

| | *Density [kg/m³]* | *$C_{10}$ [MPa]* | *$C_{20}$ [MPa]* | *$C_{30}$ [MPa]* | *$C_{40}$ [MPa]* | *$C_{50}$ [MPa]* | *$C_{60}$ [MPa]* |
|---|---|---|---|---|---|---|---|
| *Intima* | 1120 | 6.79E-03 | 5.40E-01 | -1.11 | 10.65 | -7.27 | 1.63 |
| *Media* | 1120 | 6.52E-03 | 4.89E-02 | 9.26E-03 | 0.76 | -0.43 | 8.69E-02 |
| *Adventitia* | 1120 | 8.27E-03 | 1.20E-02 | 5.20E-01 | -5.63 | 21.44 | 0.00 |

| | *Density [kg/m³]* | *E [MPa]* | *Poisson's coefficient [-]* | *Yield stress [MPa]* | *Ultimate stress [MPa]* | *Ultimate deformation [-]* |
|---|---|---|---|---|---|---|
| *Stent* | 8000 | 233E+03 | 0.35 | 414 | 933 | 45% |
| *Balloon* | 1000 | Variable (250-850) | 0.45 | - | - | - |

For the computational fluid dynamics model, transient hemodynamic analysis was performed on the final geometrical configuration from the FEA simulation using CFD with ANSYS CFX software (ANSYS, Inc., Canonsburg, PA, USA). The model included extensions at the inlet and outlets to ensure fully developed flow conditions[3]. After a sensitivity analysis, the fluid domain was meshed with tetrahedral elements, using smaller elements around the stented region (0.014 mm) to accurately capture local flow dynamics influenced by complex stent geometries (Figure 1).

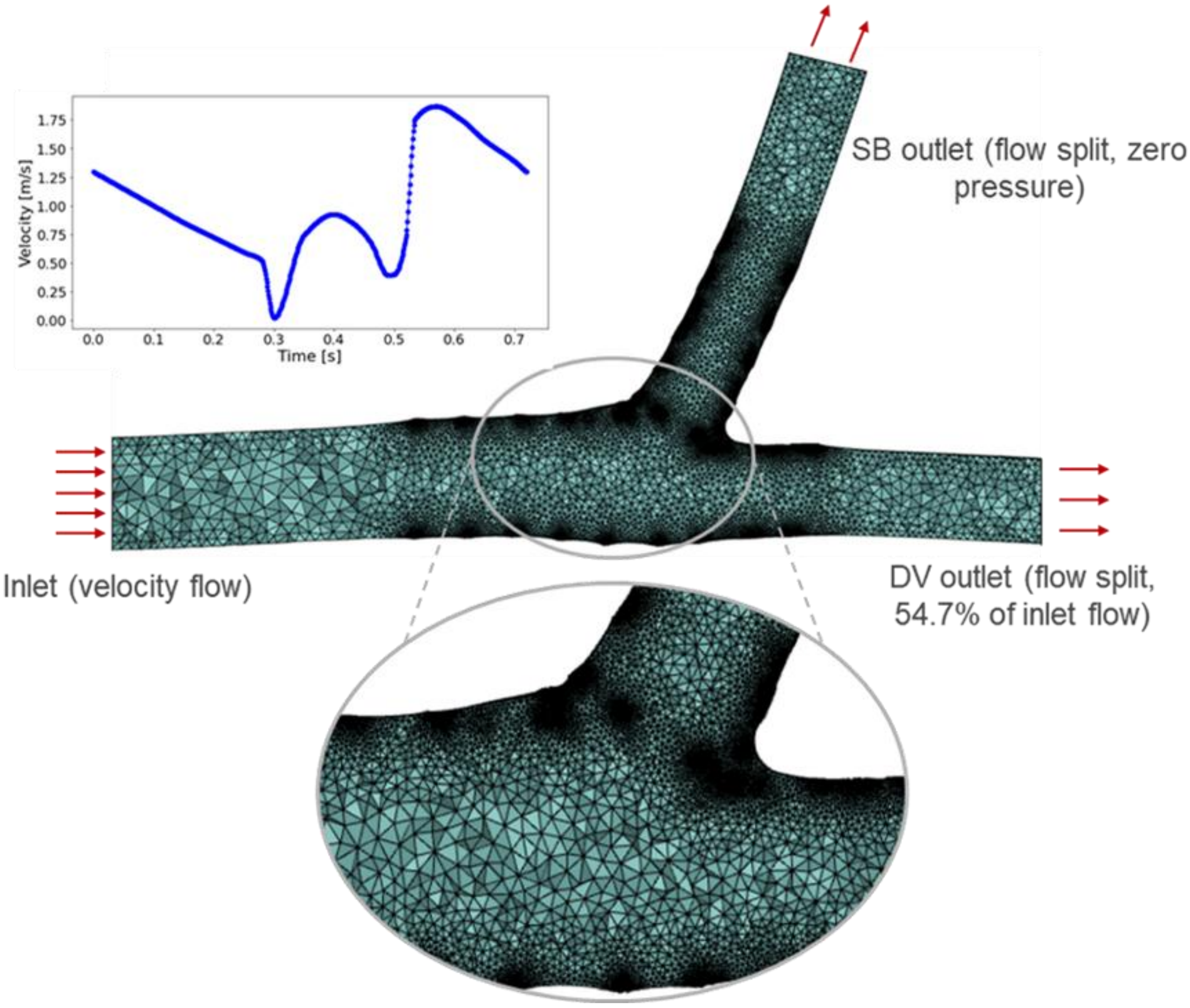

**Figure 8** – Computational fluid dynamics simulation setup for hemodynamic analysis of the coronary bifurcation model. Tetrahedral elements were used, with smaller elements concentrated in the stented bifurcation region to accurately capture complex geometries, including overlapping stent structures (magnified view at the bottom). Boundary conditions applied were a pulsatile velocity inlet representing physiological coronary flow, a flow-split outlet at the DV, and a zero-pressure condition imposed at the SB outlet. *DV: Distal Vessel; SB: Side Branch.*

The blood was modeled as an incompressible fluid with non-Newtonian viscosity using the Carreau-Yasuda model[3,4]. A parabolic pulsatile blood flow profile was applied at the MV inlet using an allometric scaling law to scale a standard velocity waveform[5], with flow rate calculated as follows[6]:

$$Q_{LM} = 1.43 {d_{LM}}^{2.55}$$

where $Q_{LM}$ is the cycle-averaged flow rate, and $d_{LM}$ is the mean diameter of MV. A flow-split outflow strategy was applied to determine the flow rate at each outlet[6]:

$$\frac{q_{LAD}}{q_{LCx}} = \left(\frac{d_{LAD}}{d_{LCx}}\right)^{2.27}$$

where $q_{LAD}$ and $q_{LCx}$ are the flow rate, and $d_{LAD}$ and $d_{LCx}$ are the mean diameters of DV and SB, respectively. Flow was split as 54.7% through DV and 45.3% through SB. The calculated flow rate was prescribed at the DV outlet, while a zero-pressure boundary condition was applied at the SB outlet to avoid over-constraint of the simulation model. The artery wall was assumed to be rigid with a standard no-slip condition applied[7]. For each model, a steady-state solution was used as the initiation condition for the transient simulation of four cardiac cycles. The results from the fourth cycle were used for analysis to minimize the transient start-up effects. TAESS was computed as follows:

$$TAESS = \frac{1}{T}\int_0^T \left|\boldsymbol{n}\,.\,\vec{\vec{\boldsymbol{\tau}}}_{\boldsymbol{ij}}\right| dt$$

where T is the cardiac cycle duration and $\vec{\vec{\boldsymbol{\tau}}}_{\boldsymbol{ij}}$ is the instantaneous wall shear stress vector. Shear rates were also evaluated to estimate thrombogenic risk, as values exceeding 1000 $s^{-1}$ are known to promote platelet activation[8]. A shear rate burden index was defined as the product of (1) the blood volume where the instantaneous shear strain rate exceeded 1000 $s^{-1}$ (extension), and (2) the average shear strain rate within that volume (intensity), capturing both the spatial and hemodynamic severity of pro-thrombotic regions.

$$Shear\ Rate\ Burden\ Index = Volume_{>1000\ s^{-1}} * Average\ Shear\ Rate_{>1000\ s^{-1}}$$

**References**

1. Gastaldi D, Morlacchi S, Nichetti R, Capelli C, Dubini G, Petrini L, Migliavacca F. Modelling of the provisional side-branch stenting approach for the treatment of atherosclerotic coronary bifurcations: effects of stent positioning. *Biomechanics and Modeling in Mechanobiology*. 2010;9:551-561. doi: 10.1007/s10237-010-0196-8
2. Morlacchi S, Chiastra C, Gastaldi D, Pennati G, Dubini G, Migliavacca F. Sequential Structural and Fluid Dynamic Numerical Simulations of a Stented Bifurcated Coronary Artery. *Journal of Biomechanical Engineering*. 2011;133. doi: 10.1115/1.4005476
3. Shen C, Gharleghi R, Li DD, Stevens M, Dokos S, Beier S. Secondary flow in bifurcations - Important effects of curvature, bifurcation angle and stents. *J Biomech*. 2021;129:110755. doi: 10.1016/j.jbiomech.2021.110755
4. Razavi A, Shirani E, Sadeghi MR. Numerical simulation of blood pulsatile flow in a stenosed carotid artery using different rheological models. *Journal of Biomechanics*. 2011;44:2021-2030. doi: https://doi.org/10.1016/j.jbiomech.2011.04.023
5. Nichols WW, O'Rourke M, Edelman ER, Vlachopoulos C. *McDonald’s blood flow in arteries: theoretical, experimental and clinical principles*. CRC press; 2022.
6. van der Giessen AG, Groen HC, Doriot P-A, de Feyter PJ, van der Steen AFW, van de Vosse FN, Wentzel JJ, Gijsen FJH. The influence of boundary conditions on wall shear stress distribution in patients specific coronary trees. *Journal of Biomechanics*. 2011;44:1089-1095. doi: https://doi.org/10.1016/j.jbiomech.2011.01.036

7. Eslami P, Tran J, Jin Z, Karady J, Sotoodeh R, Lu MT, Hoffmann U, Marsden A. Effect of Wall Elasticity on Hemodynamics and Wall Shear Stress in Patient-Specific Simulations in the Coronary Arteries. *Journal of Biomechanical Engineering*. 2019;142. doi: 10.1115/1.4043722

8. Foin N, Gutiérrez-Chico JL, Nakatani S, Torii R, Bourantas CV, Sen S, Nijjer S, Petraco R, Kousera C, Ghione M, et al. Incomplete Stent Apposition Causes High Shear Flow Disturbances and Delay in Neointimal Coverage as a Function of Strut to Wall Detachment Distance. *Circulation: Cardiovascular Interventions*. 2014;7:180-189. doi: doi:10.1161/CIRCINTERVENTIONS.113.000931

**Supporting Information 2 –** Mechanical and hemodynamic results for Orsiro and Xience Sierra after provisional side branch and double kissing crush stenting. *KBI: Kissing Balloon Inflation; MV: Main Vessel; OSI: Oscillatory Shear Index; SB: Side Branch; TAESS: Time-Averaged Wall Shear Stress.*

| | ***Provisional Side Branch (PSB)*** | | | ***Double Kissing Crush (DKC)*** | | |
|---|---|---|---|---|---|---|
| ***Stent Deployment*** | Orsiro | Xience Sierra | Difference (%) | Orsiro | Xience Sierra | Difference (%) |
| MV malapposition (%) | 13.0 | 9.6 | 3.4 | 19.1 | 15.5 | 3.6 |
| SB malapposition (%) | - | - | - | 10.3 | 7.1 | 3.1 |
| SB ostium clearance (%) | 76.9 | 64.1 | 12.8 | 52.2 | 51.7 | 0.5 |
| Max artery stress @ end [kPa] | 181.0 | 152.4 | 18.8 | 562.9 | 533.0 | 5.6 |
| Average artery stress @ end [kPa] | 22.8 | 20.0 | 13.9 | 35.0 | 33.3 | 5.0 |
| % volume with stress > 10 kPa @ end | 8.4 | 9.6 | 1.2 | 16.2 | 16.0 | 0.2 |
| Average artery stress @ 2nd KBI [kPa] | 26.6 | 24.3 | 9.6 | 42.8 | 36.3 | 17.9 |
| Max MV stent stress @ 2nd POT [MPa] | 875 | 789 | 10.9 | 871 | 884 | -1.5 |
| Max SB stent stress @ crush [MPa] | - | - | - | 712 | 806 | -11.7 |

| | *Provisional Side Branch (PSB)* | | | *Double Kissing Crush (DKC)* | | |
|---|---|---|---|---|---|---|
| ***Blood Flow*** | Orsiro | Xience Sierra | Difference (%) | Orsiro | Xience Sierra | Difference (%) |
| TAESS [Pa] | 0.52 | 0.48 | 7.5 | 0.54 | 0.47 | 13.6 |
| % area with TAESS < 0.4 Pa | 30.3 | 33.6 | -3.3 | 28.2 | 36.3 | -8.1 |
| Area with TAESS < 0.4 Pa [$mm^2$] | 36.0 | 39.1 | -7.8 | 49.0 | 58.3 | -15.9 |
| Volume with shear rate > 1000 $s^{-1}$ [$mm^3$] | 0.056 | 0.075 | -25.7 | 0.096 | 0.084 | 14.9 |
| Average shear rate > 1000 $s^{-1}$ [$s^{-1}$] | 1424 | 1513 | -5.9 | 1551 | 1527 | 1.6 |
| Shear rate burden index [$mm^3 s^{-1}$] | 79.4 | 113.5 | -30.1 | 149.7 | 128.2 | 16.7 |
| OSI [-] | 0.018 | 0.019 | -2.2 | 0.016 | 0.019 | -16.6 |